\documentclass[prc,aps,nofootinbib,showkeys,twocolumn,superscriptaddress]{revtex4}
\usepackage{epsfig}
\usepackage{graphicx}
\usepackage{amssymb}
\usepackage{color}
\usepackage{upgreek}
\usepackage{mathtools}
\usepackage[dvipsnames]{xcolor}
\usepackage{comment}
\usepackage{soul}

\usepackage{braket}

\begin{document}

\title{Construction of continuous collective energy landscapes for large amplitude nuclear many-body problems}

\author{Paul Carpentier} 
\affiliation{CEA, DAM, DIF, 91297 Arpajon, France}
\affiliation{Université Paris-Saclay, CEA, Laboratoire Matière en Conditions Extrêmes, 91680 Bruyères-le-Châtel, France}
\author{Nathalie Pillet} 
\email{nathalie.pillet@cea.fr}
\affiliation{CEA, DAM, DIF, 91297 Arpajon, France}
\affiliation{Université Paris-Saclay, CEA, Laboratoire Matière en Conditions Extrêmes, 91680 Bruyères-le-Châtel, France}
\author{Denis Lacroix }
\email{lacroix@ijclab.in2p3.fr}
\affiliation{Universit\'e Paris-Saclay, CNRS/IN2P3, IJCLab, 91405 Orsay, France}

\author{No\"el Dubray}  
\affiliation{CEA, DAM, DIF, 91297 Arpajon, France}
\affiliation{Université Paris-Saclay, CEA, Laboratoire Matière en Conditions Extrêmes, 91680 Bruyères-le-Châtel, France}

\author{David Regnier}
\affiliation{CEA, DAM, DIF, 91297 Arpajon, France}
\affiliation{Université Paris-Saclay, CEA, Laboratoire Matière en Conditions Extrêmes, 91680 Bruyères-le-Châtel, France}

\date{\today}
\begin{abstract}
Several protocols are proposed to build continuous energy surfaces of many-body quantum systems, regarding both energy and states. 
The standard variational principle is augmented with constraints on state overlap, ensuring arbitrary precision on continuity. 
As an illustration, the lowest energy and excited state paths relevant for the $^{240}$Pu asymmetric fission are studied. The scission is clearly signed, with a neutron excess in the neck, the ultimate glue before its rupture. 
Our approach can potentially connect any couple of Hilbert space states, which opens up new horizons for various applications.
\end{abstract}


\maketitle
Energy landscapes in a physical problem provide crucial information on the static and/or dynamic properties 
of the system under interest. In complex quantum systems, like interacting particles, due to the factorial 
growing of the number of degrees of freedom (DOFs) with the number of particles, such potential energy surfaces (PESs) can only be 
obtained in a rather low dimension, with a few selected, physically guided, collective DOFs. Typically, 
in atomic nuclei, these DOFs reflect the deformation of the quantum nuclear droplet. 

A recurrent difficulty observed in complex systems while calculating PESs is the occurrence 
of many discontinuities \cite{ref1,ref2,ref3,ref4,Dub12,Reg19,ref5,Zde21,ref6,ref7} either in the energies or in the overlap between adjacent states, or in both. Various origins of these discontinuities have been identified. 
For instance, when building up PESs from a variational principle, one might suddenly jump from one state to
a very different state by spontaneous symmetry breaking. Even without any discontinuity in energy, the PES
might connect two different energy landscapes that could not be dissociated when focusing solely on a few DOFs.
This leads to the difficulty that even an infinitesimal change in the selected DOFs can lead to orthogonality between many-body states.

These difficulties are fingerprints of the complexity of many-body systems, whose PESs should be built as a 
multidimensional problem with a large set of dimensions and whose evolution might reflect the existence of 
many local, almost degenerated minima. Having discontinuities and/or jumps between completely different states 
is a severe difficulty when one tries to obtain a fully quantum description of the dynamical 
properties of a many-body system, for instance when performing 
the evolution of the collective wave function, as is done in the time-dependent generator 
coordinate method \cite{Berg1,Berg,Hill1,Hill2,Gout}. Also, from first-principle arguments based on the continuity equation, an evolving quantum system should have a continuous evolution of its local density 
and, in no way, can jump instantaneously from one shape to another.  
 
A possible solution to these problems is to increase the number of relevant DOFs with the limitation that the solution of a Schr\"odinger
equation can hardly be found in more than three dimensions. Another solution is to find a practical way to get around the problem while keeping the collective 
space dimension tractable for further post-processing.  
Solving the latter problem is a milestone for the future applicability of most of the adiabatic methods, for instance applied to the problem of fission \cite{Sch16,Sch22}. 
Not surprisingly, several attempts to solve this problem in nuclear physics have been made over the years \cite{Sca19,Lau22,Las23}. 

Lastly, we would like to stress that the situation becomes even more critical when implementing non-adiabatic 
effects by including PESs associated not only with lowest energy states but also excited states \cite{Die10,Ber11,You19,Ber22,Zha22}. This problem was the original motivation of the present work. The solution we give here can be implemented in state-of-the-art nuclear density functional models and solve the problem of discontinuities for various low-lying states, including the lowest energy ones, without increasing the dimension of the collective space.  

Intending to build a smooth and continuous energy landscape both in energy and trial wave functions, we firstly reduced the problem to its simplest form:  {\it Given two many-body states $A$ and $B$, is it possible to find a continuous path between them while minimizing the total binding energy using specific variational techniques?} This problem transforms into finding a set of states that will form a path from $A$ to $B$, where we can gradually decrease the kernel to $A$ and increase the kernel to $B$. The measure of the kernel between the states $A$ and $B$ will be made using the metric $d(A,B) = {\rm Tr}(\hat D_A \hat D_B)$ where $\hat D_{A/B}$ are the density matrices of the two states. 
In this process, some states ($A$) will be repulsive since we want to escape from them, while others ($B$) will have an attractive role. Those states will be referred to as repulsors or attractors, respectively.
This simple example can be extended to a more general situation. Assume that we already have a set of states $\{C_k \}_{k=1,\cdots, K}$, some of these states being attractors and some repulsors, and we want to build a new state, using for instance a variational approach to minimize the total 
energy. This problem can be replaced by a modified variational principle, adding a set of Lagrange multipliers to control the kernel between the new state and the different states 
$C_k$.
The strategy we propose is to apply the following generalized variational principle:
\begin{eqnarray}
        \delta {\rm Tr} \left\{ \hat D \left( \hat H_c 
        - \sum_k  \beta_k [ \hat D_k  - d_k  ] - E 
        \right)
        \right\}
         = 0,\label{eq:variational}
\end{eqnarray}
where $\hat H_c$ is the Hamiltonian $\hat H$ augmented with some constraints:
\begin{eqnarray}
    \hat H_c \equiv \hat H - \sum_\alpha \lambda_\alpha [ \hat Q_\alpha - q_\alpha ].
\end{eqnarray}
$\{\hat Q_\alpha\}_{\alpha = 1,d}$ is a set of constraints, for instance the shape, the particle number\ldots of the quantum droplet. The novel aspect here is the addition of a new set of constraints on the system kernel with the configurations, through the constraint of ${\rm Tr}(\hat D \hat D_k)$. Here, $\hat D_k\equiv | C_k \rangle \langle C_k |$ are new operators
added to control the kernel of $\hat D$ with the pre-defined configurations $\vert C_k \rangle$. The Lagrange multipliers are adjusted such that, at the minimum, we have:
\begin{eqnarray}
\langle  \hat Q_\alpha \rangle = q_\alpha, ~~ \langle \hat D_k \rangle = {\rm Tr}(\hat D \hat D_k) = d_k ,\label{eq:constraint}
\end{eqnarray}
where we use the short-hand notations 
$\langle . \rangle = {\rm Tr}( \hat D .)$.
Although this method can be used in more general situations like mixed states, we will only consider here the case of pure states for which $\hat D = | \Psi \rangle \langle \Psi |$.
We note that $\langle \hat D_k \rangle$ is nothing but the kernel of the system with the configurations $\{ \vert C_k \rangle \}$. 
Eventually, $d_k$ can be set to 0 (full repulsor) or 1 (full attractor).

\begin{figure*}[htbp]
\begin{center}
\includegraphics[width =0.8\linewidth]{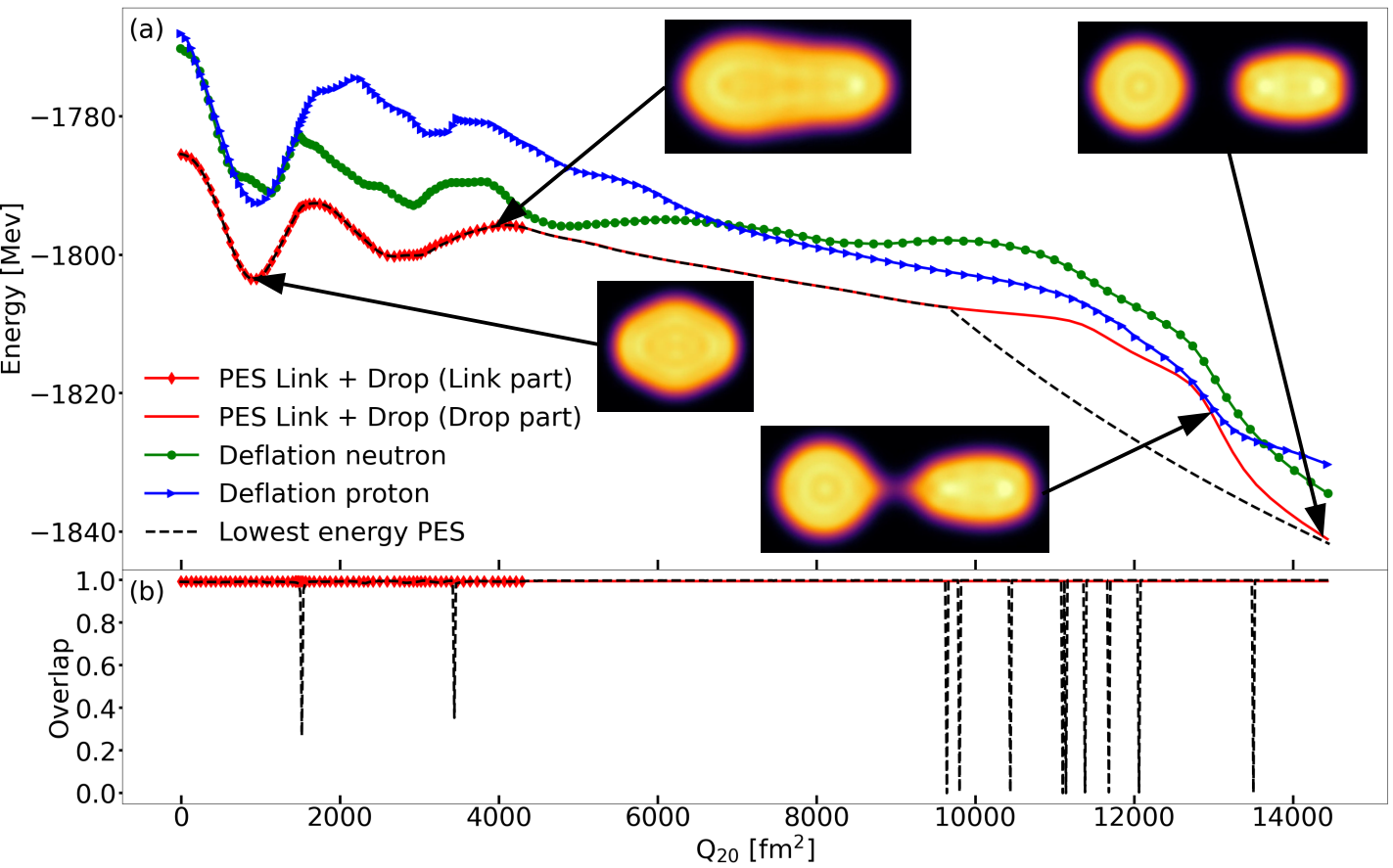}  
\end{center}
\caption{Panel (a): some PESs  (red, green, and blue curves) obtained with the different variational protocols based on Eq.(\ref{eq:variational}). They have been calculated for $^{240}$Pu as a function of the mass quadrupole moment $\langle Q_{20}\rangle$, with a continuity $d=0.990$. 
For comparison, the black dashed line corresponds to the solutions obtained with conventional minimization techniques (adiabatic limit). 
All PESs present a continuous evolution of the total binding energy. 
Note that four times more calculations have been done than those represented here by symbols for clarity reasons.
The entire red solid line has been obtained using the \textit{Link} and \textit{Drop methods}. More precisely, solutions corresponding to red diamonds are based on the \textit{Link method}, while the solutions forming the continuous red curve result from the \textit{Drop method}. After following the adiabatic path, the {\it Drop method} continuously connects to the usual Coulomb PES at $\langle Q_{20}\rangle > 14. 10^{3}$ fm$^2$. 
The different insets show spatial and local 1-body total nuclear density slices for some striking states.
The \textit{Deflation method} has been used to produce a neutron (proton) excited PES in green (blue), imposing orthogonality with the corresponding red HFB solution while ensuring the same $d$ value. Note that we have calculated ten excited states with the \textit{Deflation method} and have obtained very stable results (not shown here).
Panel (b): Overlap between each state and its adjacent state for the red 
and black curves. We can see some isolated single low values of the overlap for the black curve, corresponding to discontinuities as discussed in the introduction. 
The \textit{Link+Drop method} allows a constant high value of the overlap shown here for the red curve.
}
\label{fig:PES}
\end{figure*}
We mention that the \textit{Deflation method}, known as a general method to solve eigenvalue problems \cite{Saa03} was employed recently in quantum chemistry \cite{Hig19} and used to calculate excited states for the nuclear many-body problem \cite{Gro23,Bea23}, is a special case of the more general method we develop here. The variational principle (\ref{eq:variational}) turns out to be a versatile tool that can solve the various difficulties in constructing smooth PESs relevant to the description of atomic nuclei. 
Examples of the approach to PESs relevant to nuclear physics with a focus on the fission process are given below. 
Noteworthy, all the applications below make connection between quasiparticle (QP) vacua for which the Thouless theorem \cite{Rin80,Bla86} simplifies the variation of the constraints based on the overlaps. Indeed, starting from a QP state $| \Psi_0 \rangle$ associated with a set of quasiparticle operators $\{ \beta^\dagger_\lambda \}_{\lambda=1, \Omega}$, any surrounding QP state can be parameterized as $|\Psi({\bf Z})\rangle = e^{-\sum_{\lambda \lambda'} Z_{\lambda \lambda'} \beta^\dagger_\lambda \beta^\dagger_{\lambda'}}| \Psi_0 \rangle$. Then, the state variation reduces to the variations of the ${\bf Z}$ matrix elements and of the overlap $\langle \Psi({\bf Z}) | C \rangle$, where $|C \rangle$ is a fixed QP state:
\begin{eqnarray}
\left. \partial_{Z^*_{\lambda \lambda'} }\langle \Psi({\bf Z}) | C \rangle \right|_{{\bf Z} = 0} &=& \langle \Psi_0 \vert \beta_{\lambda} \beta_{\lambda'} \vert C \rangle,
\end{eqnarray}    
which can be evaluated numerically using standard techniques for quasiparticle states, as shown in \cite{Rob24}.

Three protocols based on the minimization of Eq.(\ref{eq:variational}) are developed below. 
The first one is called hereafter {\it Link method}.  Its goal is to link/connect two Hartree-Fock-Bogolyubov (HFB) states ($\vert A \rangle $ and $\vert B \rangle$) through a set $\vert C_i \rangle$ of HFB states iteratively, so that adjacent states present a given kernel value.
 Ultimately, the set of states $\{\vert C_{i} \rangle \}_{i=0,\dots,M}$ forms a continuous path in the many-body space from
$\vert A \rangle$ to $\vert B \rangle$. The key parameter to generate the path is a constant $0<d<1$ that fixes the squared overlap $d = |\langle C_i | C_{i-1}\rangle|^2$. 
The initial state is fixed so that $\vert C_{0}\rangle= \vert A\rangle$.
The variational principle is minimized to seek for a new state $| C_1 \rangle$ with the constraint $|\langle C_0 | C_{1}\rangle|^2 = d$ and $|\langle C_1 | B \rangle|^2$ is maximized.
Note that eventual constraints on multipole moments are not imposed, allowing the state to explore 
various shapes. The center of mass position and particle numbers are constrained during the 
minimization. At the end of this step, a state $\vert C_1\rangle$ is obtained. The procedure is then iterated.
At the i-th iteration, starting from the state $\vert C_{i} \rangle$, a new state $\vert C_{i+1} \rangle$ is generated, imposing the overlap with the previous state to be $d$ while maximizing 
the overlap with the state $\vert B \rangle$. The iteration stops when the overlap between the last generated state and $\vert B \rangle$ is greater or equal
to $d$. This method allows us to obtain low-energy PESs that are, in practice, very close to 
the adiabatic limit usually employed in nuclear physics while being continuous both in energy 
and overlap. In practice, one can choose a value $\vert d-1 \vert \le \epsilon$, 
requiring $\epsilon \ll 1$ to give direct control of the continuity in the overlap of adjacent 
states along the path. Decreasing $\epsilon$ leads to an increase in the
number of trial states that are used to build the PESs.

Even if the {\it Link method} works well to connect two HFB states, the requirement of having both a starting and a final states is no longer fulfilled when it comes to describing situations wherein final configurations are not known {\it a priori}, as for the scission process. 
The {\it Drop method} has been developed to tackle this issue. 
It creates an adiabatic and continuous path from a starting configuration, only following an energy descent.
This method is "goal-free" and thus enables us to adequately describe processes such as the scission one. It starts with an HFB state $\vert A \rangle = \vert C_0 \rangle$ and finds the state $\vert C_1 \rangle$ minimizing the energy while ensuring $|\langle C_0 | C_{1}\rangle|^2 = d$. 
This procedure is then iterated from $\vert C_i \rangle$ to find $\vert C_{i+1} \rangle$. 
It stops after a given number of iterations or whenever the energy of $\vert C_{i+1} \rangle$ exceeds that of $\vert C_{i} \rangle$. 
While the {\it Link method} allows to obtain continuous low-energy PESs from discontinuous ones, staying very close to the adiabatic limit usually employed in nuclear physics, the {\it Drop method}, where the constraint on the final target is relaxed, generates paths that wouldn’t have been possible to describe without it. 

The {\it Drop method} is efficient on the descent to scission when usual approaches using multipole moments struggle to describe the nucleus continuously just before scission and fail most of the time to obtain reliably separated fragments. 
Those difficulties are systematically observed in such studies, and tentative solutions have been proposed using additional geometrical constraints \cite{You11,War12,Mar21}, but without controlling the continuity of the states. 
Continuously crossing the scission is crucial to make contact with experimental observations since, after separation, daughter nuclei essentially only encounter the Coulomb boost with eventual in-flight post-equilibrium emissions.
Methods to extract the PES were also proposed based on the nuclear Time-Dependent Density Functional Theory (TDDFT). TDDFT has the advantage of ensuring continuity in the trial state vector and can also go through the scission region. However, specific methods should be developed to separate static and dynamical effects in the PES  \cite{Uma10,Tan15}. Additionally, the path followed in TDDFT is sometimes too diabatic, preventing the description from the saddle point \cite{God15, God16}, and explores restricted regions in the collective space \cite{Bul19}.

PESs obtained by combining the {\it Link} and {\it Drop methods} are shown in Fig.\ref{fig:PES}, panel (a). The D1S Gogny interaction has been used \cite{Berg,Berg1,Gogny1,Gogny2,Gogny3} as well as a new HFB solver whose solutions are expanded on an axial two-center harmonic oscillator basis allowing the breaking of parity symmetry \cite{HFB3}. The standard technique to generate adiabatic PESs (black dashed line) leads to a continuous energy evolution but significant kernel jumps in adjacent state vectors obtained by energy minimization. For instance, the second sudden jump observed in the overlap shown in Fig.\ref{fig:PES}, panel (b), signs the quantum phase transition from symmetric to asymmetric nuclear shapes due to the spontaneous breaking of the parity symmetry,
the first discontinuity at large quadrupole moment $Q_{20}$ signs the Coulomb-fission valleys crossing, and the following ones highlight the complex structure of the Coulomb valley.
The {\it Link method} allows the generation of a PES very close to the adiabatic one while ensuring an overlap between adjacent states that is precisely controlled using a high kernel value $d=0.990$. Noteworthy, using the {\it Drop method} at a large quadrupole moment $Q_{20}$ also leads to a continuous overlap while connecting without problem the PESs to the Coulomb energy surface. As far as we know, this is the first-ever continuous connection of the fission valley to the Coulomb valley using such fine quantitative control of the overlap between adjacent states. 

\begin{figure}
    \centering
    \includegraphics[width=\linewidth]{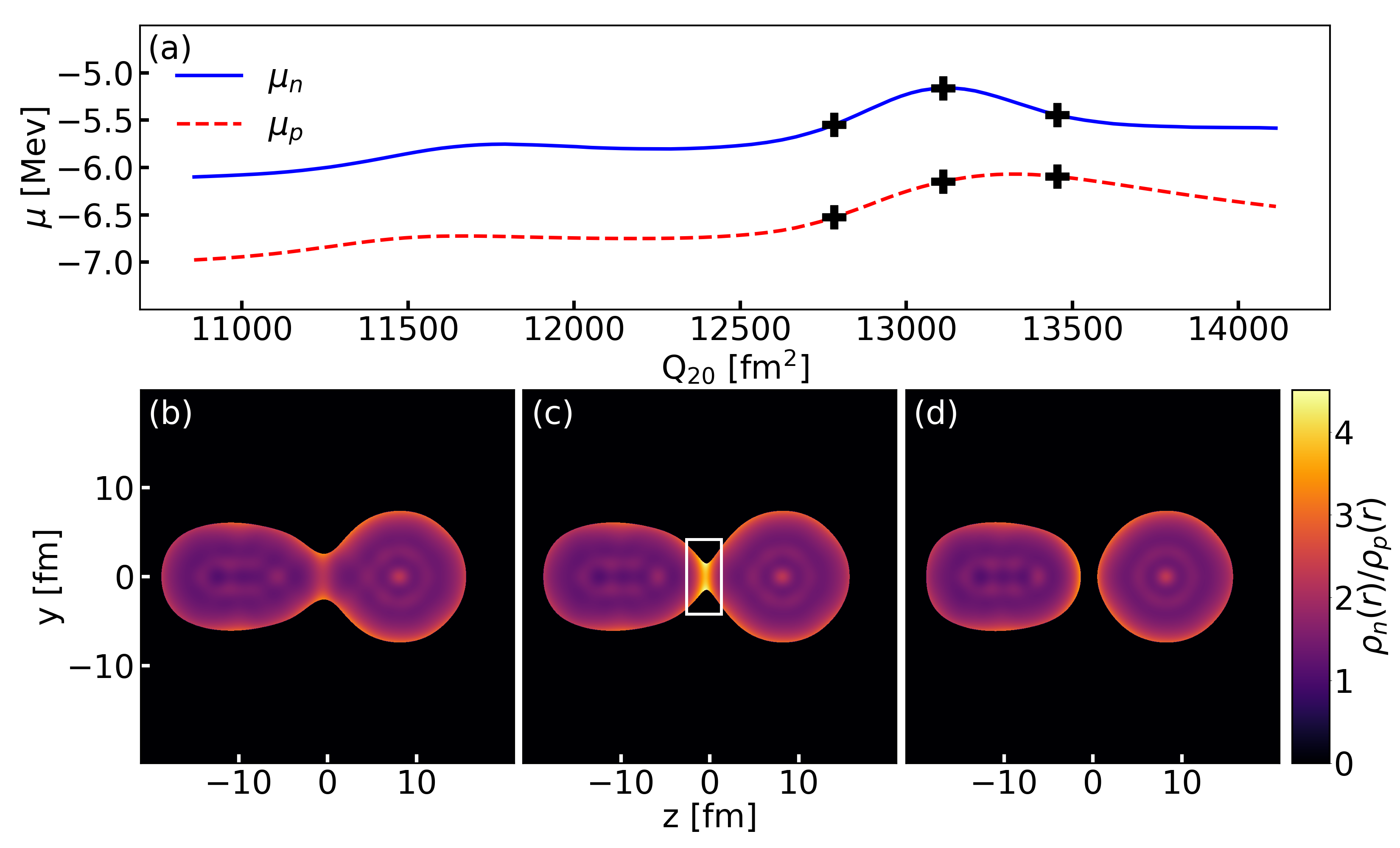}
    \caption{Panel (a): Evolution of the neutron (blue line) and proton (red dashed line) chemical potentials $\mu_{n,p}$ along the \textit{Link+Drop} path shown in Fig.\ref{fig:PES}. Panels (b-d): Ratio $\rho_n({\bf r})/\rho_p({\bf r})$ between neutron and proton 1-body local densities for a total density $\rho({\bf r})$ exceeding the threshold value $5.10^{-3}$ fm$^{-3}$. The (b) to (d) states are indicated in panel (a) by black crosses. The white box in panel (c) delimits the region where the neutron-to-proton fraction is maximally enhanced. 
    } 
    \label{fig:neutrondistillation}
\end{figure}
The possibility of accessing the full path to dissociation in $^{240}$Pu allowed us to uncover an interesting phenomenon illustrated in Fig \ref{fig:neutrondistillation}, panel (a), which 
shows the neutron and proton chemical potentials, $\mu_n$ and $\mu_p$, along the {\it Link}+{\it Drop} path. 
As the fragments separate, we see that neutrons are less bound than protons with 
a sudden increase of $\mu_{n,p}$ that signs the neck rupture. Just before the rupture, an excess of neutrons appears in the neck. This is illustrated in Fig.\ref{fig:neutrondistillation}, panels (b-d). Here, the ratio between the local neutron and proton 1-body densities reaches values three times larger than the ratio $N/Z = 1.55$.  
This enrichment is due to a single neutron quasiparticle state strongly localized at the neck before the scission that serves as the ultimate glue between the two daughter nuclei before the separation. This phenomenon looks very similar to the electrons that bind molecules or the generalized version of the Ikeda diagram introduced for clustering classification in neutron-rich nuclei \cite{von06}, excepted that for fission, this configuration is an intermediate step towards the final separation. Finally, the mean total kinetic energy, estimated by simply accounting for the Coulomb re-acceleration at the neck rupture assumed to occur at the maximum of $\mu_{n}$ (second cross in Fig.\ref{fig:neutrondistillation}, panel (a)), is found to be 175.4 MeV. This value is 
very close to the 177.7 MeV measured experimentally \cite{Rei71,Bou91}.

A third protocol based on Eq.(\ref{eq:variational}), called hereafter {\it Deflation}, is proposed to build continuous PESs for excited states.  
Provided that we have already constructed a set of states $\{ \vert C_k \rangle \}_{k=1,\cdots,M}$ associated with the lowest energy landscape with the {\it Link} and {\it Drop methods},
we then use these states as pure repulsors to build the first excited state PES. 
As an example relevant for fission studies, for a neutron (proton) excitation, Eq.(\ref{eq:variational}) is solved imposing that the neutron (proton) part of the HFB reference state is orthogonal to the neutron $\vert C_i \rangle$ state one ($d_i \simeq 0$) while the proton (neutron) parts of both states are required to be equal ($d_{i'} \simeq 1$).
Having the first excited state PES, the second excited state one is obtained by imposing the simultaneous orthogonality to the lowest energy and the first excited PESs, and so on and so forth. Two examples of such excited state PESs, obtained by imposing orthogonality either with the neutron or proton lowest energy wave functions, are shown in Fig.\ref{fig:PES}, panel (a). In both cases, a near-perfect orthogonality with the lowest energy states and a near-perfect continuity between adjacent excited states are reached.
The method has been tested to build up to ten successive excited states of this type without specific difficulties. More general excited states can also be generated by the {\it Deflation method}.

By adding constraints on overlaps when building collective PESs for many-body self-bound systems, we propose a set of versatile protocols based on the variational principle (\ref{eq:variational}) that ensures continuity both in energy and in the states themselves for a set of low-lying states, without increasing the number of DOFs. While PESs with continuous energy are standard, the control of continuity in states has been missing so far. It will open opportunities in applications that were difficult or even impossible until now. To quote some of them in the nuclear physics context, one can mention the possibility of performing time-dependent configuration mixing using overlap kernels without approximation. The availability of excited state PESs will also allow the inclusion of non-adiabatic effects during the evolution and the onset of internal excitation \cite{Ber11,Die10}. This is one of the paths to describe dissipative effects and quantum collective fluctuations on the same footing in a common framework. In a completely different topic, since the method also allows to connect various configurations in the Hilbert space, one can use it to address the problem of clustering in atomic nuclei and connect in a continuous way normal configurations with configurations presenting clusters, as well as cluster emission \cite{Warda2,Gir13,Zdeb2,War18,Ebr20,Mer21,von06}. \\

\noindent {\it Acknowledgement:} NP and PC would like to thank L. Robledo, R. Bernard and 
W. Younes for valuable discussions. DL thanks Y. Beaujeault-Taudi\`ere for the discussions on the \textit{Deflation method} at the early stage of this work. DL acknowledge the support of the IN2P3-AIQI project.

\end{document}